\documentstyle[psfig]{l-aa}

\def\jref#1 #2 #3 #4 {{\par\noindent \hangindent=2em \hangafter=1
      \advance \rightskip by 0em #1, {\it#2}, {\bf#3}, #4.\par}}
\def\rref#1{{\par\noindent \hangindent=2em \hangafter=1
      \advance \rightskip by 0em #1.\par}}

\def\deg{\ifmmode^{\circ}\else$^{\circ}$\fi} 
\def\min{\ifmmode^{\prime}\else$^{\prime}$\fi}
\def\sec{\ifmmode^{\prime\prime}\else$^{\prime\prime}$\fi}
\def\fmin{\ifmmode.\!\!^{\prime}\else$.\!\!^{\prime}$\fi}
\def\fsec{\ifmmode.\!\!^{\prime\prime}\else$.\!\!^{\prime\prime}$\fi}

\def\fd{\hbox{$.\!\!^{\rm d}$}}

\def\lapp{\ifmmode\stackrel{<}{_{\sim}}\else$\stackrel{<}{_{\sim}}$\fi}
\def\gapp{\ifmmode\stackrel{>}{_{\sim}}\else$\stackrel{>}{_{\sim}}$\fi}
\newcommand{\xr}{X-ray}
\newcommand{\gr}{gamma-ray}

\newcommand{\fu}{erg~cm$^{-2}$ s$^{-1}$}
\newcommand{\SAX}{{\em Beppo}SAX}
\newcommand{\B}{GRB970402}
\newcommand{\SJ}{1SAX J1450.1--6920}
\newcommand{\etal}{et~al.\ }
\newcommand{\eal}{et~al.}

\begin{document}

\title{\SAX\ observations of \B}

\author{L. Nicastro\inst{1}
\and L. Amati \inst{2,8}
\and L. A. Antonelli \inst{3}
\and R. C. Butler \inst{9}
\and E. Costa \inst{2}
\and G. Cusumano \inst{1}
\and M. Feroci \inst{2}
\and F. Frontera \inst{4,5}
\and J. Heise \inst{6}
\and K. Hurley \inst{10}
\and J. M. Muller \inst{3,6}
\and A. Owens \inst{7}
\and L. Piro \inst{2}
}

\offprints{nicastro@ifcai.pa.cnr.it}
\date{Received ........; accepted .........}
\thesaurus{011 (13.07.1; 13.25.1)}

\institute{
{Istituto di Fisica Cosmica con Applicazioni all'Informatica, CNR,
 Via U. La Malfa 153, I-90146 Palermo, Italy}
\and 
{Istituto Astrofisica Spaziale, CNR, Via Fosso del Cavaliere, I-00131 Roma,
 Italy}
\and 
{BeppoSAX Scientific Data Center, Via Corcolle 19, I-00131 Roma, Italy}
\and 
 {Istituto Tecnologie e Studio Radiazione Extraterrestre, CNR, Via Gobetti 101,
 I-40129 Bologna, Italy}
\and 
 {Dipartimento di Fisica, Universit\`a di Ferrara, Via Paradiso 11,
 I-44100 Ferrara, Italy}
\and 
{Space Research Organization in the Netherlands, Sorbonnelaan 2,
 3584 CA Utrecht, The Netherlands}
\and 
{Astrophysics Division, Space Science Department of ESA, ESTEC,
P.O. Box 299, 2200 AG Noordwijk, The Netherlands}
\and 
{Istituto Astronomico, Universit\`a degli Studi ``La Sapienza'', via Lancisi 29,
 Rome, Italy}
\and 
{Agenzia Spaziale Italiana, V.le R. Margherita 120, Roma, Italy}
\and 
{Space Sciences Laboratory, University of California, Berkeley, CA 94720--7450,
 USA}
}

\maketitle
\markboth{L. Nicastro et al.: \SAX\ observations of \B}
         {L. Nicastro et al.: \SAX\ observations of \B}

\begin{abstract}
\B\ is the fourth \gr\ burst detected by \SAX\ simultaneously in the
Gamma Ray Burst Monitor (GRBM) and one of the two Wide Field Cameras (WFCs).
A rapid pointing of the \SAX\ Narrow Field Instruments (NFIs)
8 hours after the GRB led to the identification of an unknown weak \xr\ source:
\SJ. Its position was approximately at the center of the 3 arcmin
error circle derived from the WFC image. Both the Medium Energy (MECS, 2--10
keV) and Low Energy (LECS, 0.1--10 keV) concentrators detected the source.
A follow-up observation performed 1.5 days later and lasting 54 ks showed that
the source had faded almost to, but not below the detectability threshold.
The flux decrease between the two observations was a factor $\simeq 2.5$.
\SJ\ was the second X-ray afterglow associated with a GRB.
Searches promptly started at other wavelengths (optical, IR, radio)
did not reveal any transient event within the 3 arcmin error circle.


\end{abstract}

\section{Introduction}

The long-investigated nature of Gamma Ray Bursts (GRBs) has been
mostly a matter of theoretical speculation for more than twenty years (for a
review see e.g. Fishman \& Meegan 1995).
The rapid identification and accurate ($\approx$ 1 arcmin) determination of
their positions was the only way to initiate a new era
in this field of astronomy. This became possible with the launch
of the \SAX\ satellite (Boella \etal 1997a).
Descriptions of the \SAX\ Gamma Ray Burst Monitor (GRBM, 40--700 keV),
the two coded mask Wide Field Cameras (WFCs, 1.5--26 keV)
and the quick-look analysis performed in order to localize GRB events 
are given elsewhere
(Frontera \etal 1997; Feroci \etal 1997a; Costa \etal 1998; Jager \etal 1997;
Costa \etal 1997a).

\B\ is the fourth GRB detected by one of the two \SAX\ WFCs.
The burst was also detected by Ulysses but not by BATSE.  
As for GRB970228 (Costa \etal 1997a), thanks to a prompt observation by
the \SAX\ NFIs, an \xr\ afterglow was identified, but, unlike that event,
no detection at other wavelengths has been reported for \B.

In this paper we report and discuss the \gr\ and \xr\ results of \B\ and its
afterglow.

\section{BeppoSAX detection, TOO observations and data analysis}

\begin{figure}[tb]
\psfig{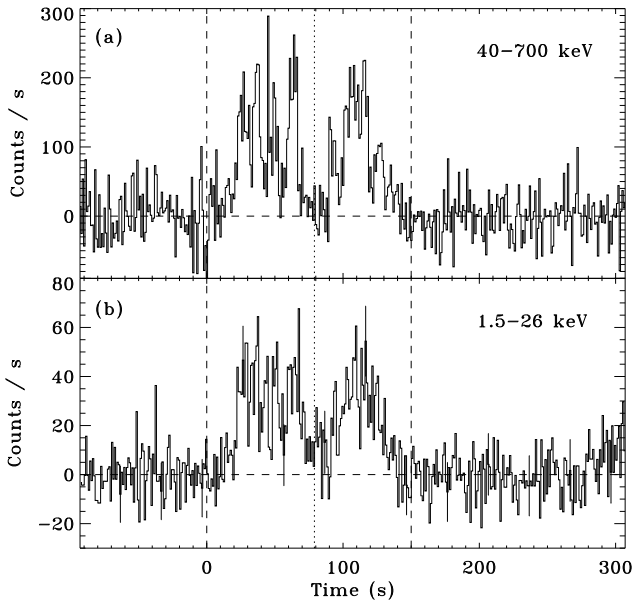}
\caption[]{Light curves of GRB970402 with 1 s time resolution.
 (a) in the PDS--GRBM
 unit 1, (b) in the WFC unit 1. Adopted start, mid and end time of the
 burst are shown.}
\label{fig:b4lc}
\end{figure}

The GRBM was triggered on April 2, 22:19:39 UT (Feroci \etal 1997b).
The 40--700 keV light curve showed a weak, structured burst with 
two major broad peaks with durations of $\sim 80$ and $\sim 70$ s respectively.
Analysis of the ratemeters from the WFC1 confirmed the GRBM results
of a burst lasting $\simeq 150$ s (see Fig. \ref{fig:b4lc}).

In the 40--700 keV energy range the peak flux was
$f_\gamma = (2.4\pm0.7)\times 10^{-7}$ \fu\ and
the fluence (over 150 s) $F_\gamma = (8.2\pm0.9)\times 10^{-6}$ erg cm$^{-2}$.
In the WFC unit 1 the 2--10 keV burst peak flux was
$f_X = (7\pm 2)\times 10^{-9}$ \fu\ and the fluence
$F_X = (4.2\pm 0.4)\times 10^{-7}$ erg cm$^{-2}$ with
the ratios $f_{X/\gamma} = 0.03$ and $F_{X/\gamma} = 0.05$.
Spectral evolution analysis using the WFC and GRBM data indicate that
all the spectra can be well fitted with a single power law. The spectral photon
index ($N_E\propto E^\alpha$) of the combined WFC + GRBM data of the
entire burst is almost constant at a value of
$-1.38\pm 0.25$. Fitting only the WFC data gives an index of $-1.31\pm0.18$
for the first peak and $-1.50\pm0.22$ for the second one.

The burst position derived from the WFC image was (equinox 2000)
R.A.=$14^{\rm h}\, 50^{\rm m}\, 16^{\rm s}$,
Decl.=$-69\deg 19\fmin9$ with an error radius of 3 arcmin
(Heise \etal 1997). Triangulation  using Ulysses and \SAX-GRBM arrival
times gives an annulus consistent with this position, but does not
reduce the size of the error circle (Hurley \etal 1997).

\begin{figure*}[tb]
\centerline{
\psfig{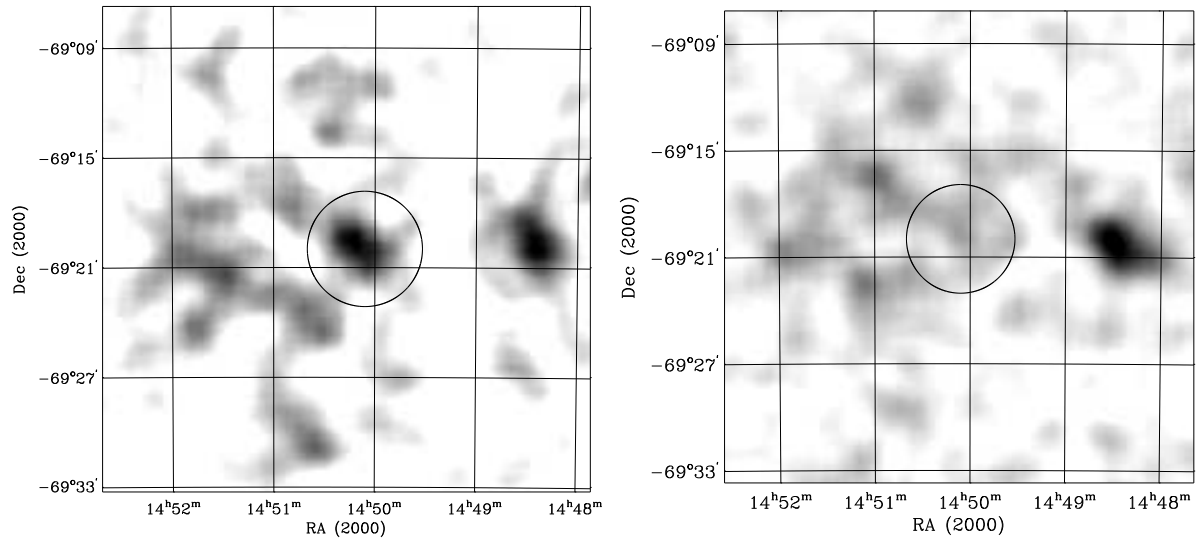}
}
\caption[]{MECS (2--10 keV) images of GRB970402 for the two \SAX\ observations:
 $\sim 8$ hr (left) and 1\fd7 (right) after the burst.
 The WFC $3'$ error circle is shown.}
\label{fig:too12}
\end{figure*}

\SAX\ performed two target of opportunity (TOO) observations of \B\ with its
NFIs. The first observation started on April 3.280 UT, $\sim 8$ hours
after the burst trigger. The total exposure time was 34.5 ks in the MECS
and 11.6 ks in the LECS.
Two sources are detected at a high statistical significance (see Fig.
\ref{fig:too12}). \SJ\ is near the
center of the WFC error circle. The second serendipitous source is $\sim 10$
arcmin west of it and will not be investigated in this paper.
\B\ seems to lie on the Eastern edge of a diffuse quasi-circular emission
region. We suggest that it is an unidentified shell-like SNR. Persistence of the
X-ray emission in the second TOO confirms this hypothesis.

Analysis of the three MECS units independently, shows that \SJ\
is only marginally detected in unit 1.
This is due to the lower sensitivity of this unit at energies below 4 keV.
Also the MECS unit 2 shows a more stable background respect to unit 3, so
it is more suitable for reconstructing the
source position (this is true in particular for weak and off-axis sources).
The source position was thus obtained using the MECS2 data only giving
(equinox 2000)
R.A.=$14^{\rm h}\, 50^{\rm m}\, 06^{\rm s}$, Decl.=$-69\deg\, 20\fmin0$
with a conservative error radius of 50 arcsec (90\% confidence level).
The position obtained with the LECS data is consistent with that of MECS2.
We computed the extraction circle giving the highest signal to noise
ratio to be 2\fmin4 for the MECS and $8'$ for the LECS.
For the spectral analysis, due to the presence of the SNR, the standard
background subtraction using blank sky maps could not be performed.
For the MECS we found that a good background estimate can be derived using
the second TOO (longer than the first one), accumulating the counts
in a circular annulus,
centered on the source position, with radii 2\fmin4--4\fmin8.
For the LECS we also used the data of the second TOO. In this case, because
of the source decay, we could accumulate the background spectrum
in an $8'$ circle centered on the source but
using only the data of the second part of the observation.

Spectral fits with a power law model for the LECS and MECS2+3 data together
were performed. The best results were obtained using an absorbed
power law. For the $N_H$ we obtain only an upper limit of $2\times 10^{22}$
cm$^{-2}$ (90\% confidence level, best value $1\times 10^{21}$cm$^{-2}$).
The photon index is $\alpha = -1.7\pm 0.6$. The reduced $\chi^2$ is 1.1
(for 12 degrees of freedom).
The average flux in the 2--10 keV MECS band is
$(2.2\pm0.6)\times10^{-13}$ \fu,
while in 0.1--2 keV band (LECS)
we observe a flux of $(1.9\pm0.7)\times10^{-13}$ \fu.

\SAX\ performed a second observation of \B\ on April 4.634, $\sim 1.7$ days
after the
burst, lasting about 65 ks. Exposure times were 54.3 ks for the MECS and
13.7 ks for the LECS.
As can be seen from Fig. \ref{fig:too12}, \SJ\ is only marginally detected
in the MECS.
One can argue that this could be due to the underlying SNR and not to the
afterglow
itself. We believe this is not the case because, splitting the observation
into three parts, we see a statistically significant count rate decrease;
in fact we use only the first 50 ks to calculate the source flux. We obtain
$f_X = (9.7\pm4.4)\times10^{-14}$ \fu\ (2--10 keV).
In the LECS (0.1--2 keV) we only can derive a $3\sigma$ upper limit of
$\sim 2\times 10^{-13}$ \fu.

\begin{figure}[tb]
\psfig{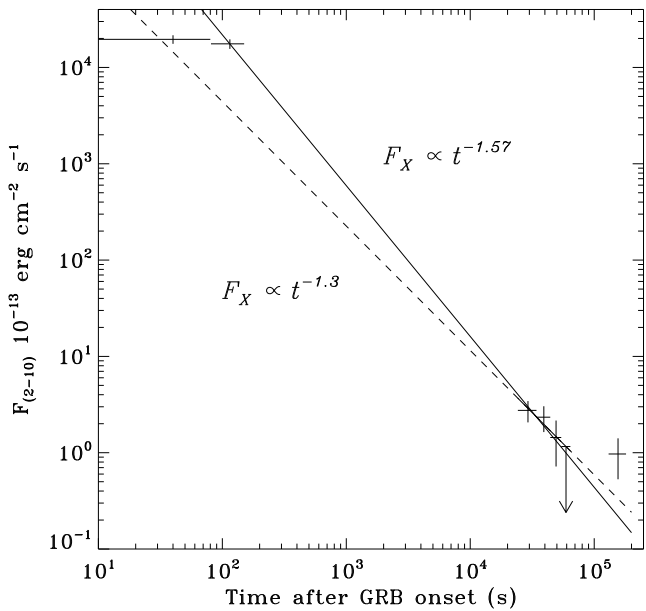}
\caption[]{WFCs and MECS light curve of GRB970402 in 2--10 keV.
 The first two bins correspond to the two peaks of the burst.
 Two possible decay laws are shown (see text).}
\label{fig:lc}
\end{figure}
The 2--10 keV light curve with $\pm 1\sigma$ statistical errors of the MECS,
together with the WFC points corresponding to the two GRB peaks,
are shown in Fig. \ref{fig:lc}. The fourth bin of the first TOO is a $3\sigma$
upper limit.
In spite of the low statistics in the two TOOs, it is apparent that
there is not a constant
flux decay; note the  possible flattening after the end of the first TOO.
We performed a least squares fit with a power law $F_t \propto t^{-\delta}$
(where $t$ time from the GRB onset) to the
first TOO points obtaining $\delta$ in the range $-1.3\div -1.6$.
Constraining the fit with the second WFC point (see discussion) and also
including the second TOO point, gives $\delta = -1.57\pm0.03$.
This last value is considerably steeper than the $-1.33$ of GRB970228
(Costa \etal 1997a) and the $-1.1$ of GRB970508 (Piro \etal 1998).

\section{Discussion and conclusions}
Optical monitoring was promptly started at various sites, but failed to detect
any transient object at a level of R=21.5 (Groot \etal 1997), V=22.5 and I=21.7
(Pedersen \etal 1997).
Moreover, two ISO (Infrared Space Observatory) observations performed
two and twelve days after the burst did not detect any variable IR object
(Castro--Tirado \etal 1998).
Radio observations were performed at frequencies between 1.5 and 8.6 GHz
using the Australia Telescope Compact Array starting  April 3,
15:12 UT and lasting until July 3. No radio source was discovered within the
X-ray afterglow error circle (Frail 1997).
While the lack of an optical detection could be attributed to a high 
Galactic extinction in this direction ($b\simeq -9^\circ$), the non-detection
in the IR and radio suggests it could more likely be due to a rapid decay law.

Including GRB970111 (Feroci et al. 1998), \B\ represents the third GRB
afterglow discovered at X-ray wavelengths. A comparison of the spectral
and temporal behaviour of this burst in the framework of the results from the
entire sample of GRBs with detected afterglows is beyond the scope of this
paper. A review of the spectral properties of all the GRBs detected by \SAX\
is presented by Frontera \etal (1998b).

\B\ has a peak flux in the GRBM and WFC bands more than one order of magnitude
lower   
than that of GRB970228, while their ratio $f_{X/\gamma} = 0.03$ is
comparable to the 0.04 of the February event (Frontera \etal 1998a) and to the
0.01 of GRB970111 (Feroci et al. 1998).
On the other hand the fluence ratio $F_{X/\gamma} = 0.05$ is 4 times smaller
than that of GRB970228 and 10 time smaller than that of GRB970508,
but similar to the 0.04 of GRB970111.
This could suggest some correlation with the presence or lack of an optical
transient (GRB970228 and GRB980508 have it, not the other two) but at least
for another burst, GRB980329, the fluence ratio is only $\simeq 0.01$
(in 't Zand \etal 1998) but an optical counterpart was detected
(Palazzi \etal 1998).
Also, in spite of the many differences in the morphology of GRB970228 and \B,
their 2--10 keV mean flux ratio $F(GRB970228)/F(\B)\simeq 15$ does not change
in the afterglow (i.e. after $\sim 12$ hours from the high energy event).

From the data of the two TOOs alone we note that the decay is not monotonic
therefore it is difficult to uniquely determine the flux decay index.
However, if we fit the first TOO data taking into account the fourth time bin
upper limit, we obtain $\delta$ in the range $-1.3\div -1.6$.
The steeper index is obtained by setting the upper limit to its formal value
decreased by $1\sigma$.
Both these values give an extrapolated flux at the time of the GRB consistent
with that measured by the WFC (see Fig. \ref{fig:lc}). This might be an
indication that the GRB and afterglow X-ray emission have a common origin
and that, as for GRB970228, the afterglow starts some tens of seconds
after the gamma-ray burst onset (Frontera \etal 1998a).
If we then add the WFC flux of the second
GRB peak to the fit we obtain a slope $\delta=-1.57\pm 0.03$,
consistent with the range found from the first TOO alone. This result
does not change adding the second TOO point to the fit.
This would be the most rapid decay so far measured.

Simple fireball models predict a flux evolution $\propto t^\delta$ with
$\delta=(3/2)(\alpha+1)$ (Tavani 1997; Wijers \etal 1997).
Considering the two decay indices $\delta=-1.3$ and $-1.57$, we obtain
$\alpha=-1.9$ ($\pm 0.4$) and $\alpha=-2.05$ ($\pm 0.05$) respectively.
These values are both in agreement with the $-1.7\pm0.6$ we find from
the spectral fitting of the first TOO. On the other hand only the first
value is fully consistent with the $-1.38\pm0.25$ of the entire
GRB event and $-1.5\pm0.2$ found in the WFC data of the second peak
($\sim 100$ s after the burst onset).
So, due to the uncertainties on the flux decay index and low statistics,
we do not see evidence of X-ray spectral evolution.

GRB970111 has a much higher flux and quite peculiar spectral evolution
compared to \B\ (Frontera \etal 1998b).
However, in addition to the small $f_{X/\gamma}$ and $F_{X/\gamma}$,
these two bursts are remarkably similar when considering their afterglow
X-ray flux of $\sim 1\times 10^{-13}$ \fu\ after $\sim 1\times 10^5$ s from
the GRB onset, the lack of an optical/IR/radio transient and the
$\delta\simeq-1.5$.
Also the two GRBs profiles have similar multi-peak structure.

Even though the X-ray flux is not intense enough for a study with good
statistical significance, the afterglow flux decay does not clearly display a
constant behaviour. Again at a low significance, this is also true
for GRB970111 and GRB970228. In the case of GRB970508, the flux was ten times
higher and allowed a clear demonstration of the departure from a smooth decay
(Piro \etal 1998) of the X-ray flux;
the same was true of ASCA observations of GRB970828 (Yoshida \etal 1997).
A fireball model with a highly radiative, relativistically expanding
shell (Cavallo \& Rees 1978; Rees \& M\'esz\'aros 1992) cannot explain
such behaviour unless, for example, a variable density for the surrounding
environment is invoked (Vietri 1997), or an asymmetric shell with temporally
evolving patchy X-ray active regions (Fenimore \etal 1998).
Future X-ray afterglows will determine whether (as well could be the case),
the non-constant fading is the rule and not the exception.

\begin{acknowledgements}
This research is supported by the Italian Space Agency (ASI) and
Consiglio Nazionale delle Ricerche (CNR). BeppoSAX is a major program
of ASI with participation of the Netherlands Agency for Aerospace
Programs (NIVR). All authors warmly thank the extraordinary teams
of the BeppoSAX Scientific Operation Center and Operation Control
Center for their enthusiastic support to the GRB program.
K. H. is grateful to the US SAX Guest Investigator program for support.
\end{acknowledgements}


\begin{thebibliography}{}
%
\bibitem{}
Boella G., \eal, 1997a, A\&AS, 122, 299
%
\bibitem{}
Boella G., \eal, 1997b, A\&AS, 122, 327
%
\bibitem{}
Castro-Tirado A., \eal, 1998, A\&A, 330, 14
%
\bibitem{}
Cavallo G., M.J. Rees, 1978, MNRAS, 183, 359
%
\bibitem{}
Costa E., \eal, 1997a, Nat, 387, 783
%
\bibitem{}
Costa E., \eal, 1997b, IAUC 6533
\bibitem{}
Costa E., \eal, 1998, Adv. Sp. Res., in press.
%
\bibitem{}
Fenimore E.E., \eal, 1998, ApJ (submitted, astro-ph/9802200)
%
\bibitem{}
Feroci M., \eal, 1997a, Proc. SPIE 3114, 186
%
\bibitem{}
Feroci M., \eal, 1997b, IAUC 6610
%
\bibitem{}
Feroci M., \eal, 1998, A\&A, in press.
%
\bibitem{}
Fishman G.J., Meegan C.A., 1995, ARA\&A, 33, 415
%
\bibitem{}
Frail D., WEB page: \\
 http://www.nrao.edu/$\sim$dfrail/grbvla/grb.html
%
\bibitem{}
Frontera F., \eal, 1997, A\&AS, 122, 357
%
\bibitem{}
Frontera F., \eal, 1998a, ApJ, 493, L67
%
\bibitem{}
Frontera F., \eal, 1998b, ApJ (in preparation)
%
\bibitem{}
Groot P., \eal, 1997, IAUC 6616
%
\bibitem{}
Heise J., \eal, 1997, IAUC 6610
%
\bibitem{}
Hurley K., \eal, 1997. In: Proc. The Active X-ray Sky, 21--24
 Oct. 1997, Rome (I) (in press)
\bibitem{}
in 't Zand J.J.M., \eal, 1998, ApJ, submitted
\bibitem{}
Jager R. \eal, 1997, A\&AS, 125, 557
%
\bibitem{}
Katz J.I., Piran T., 1997, ApJ, 490, 772
%
\bibitem{}
Klebesadel R.W., \eal, 1973, ApJ, 182, L85
%
\bibitem{}
Palazzi E., \eal, 1998, A\&A, in press
\bibitem{}
Parmar A., \eal, 1997, A\&AS, 122, 309
%
\bibitem{}
Pedersen H., \eal, 1997, IAUC 6628
%
\bibitem{}
Piro L., \eal, 1997, IAUC 6617
%
\bibitem{}
Piro L., \eal, 1998, A\&A 331, L41 (astro-ph/9710355)
%
\bibitem{}
Rees M.J., M\'esz\'aros P., 1992, MNRAS, 258, 41
%
\bibitem{}
Tavani M., 1997, ApJ, 483, L87
%
\bibitem{}
van Paradijs J., \eal, 1997, Nat, 386, 686
%
\bibitem{}
Vietri M., 1997, ApJ, (submitted, astro-ph/9706060)
%
\bibitem{}
Wijers R.A.M., Rees M.J, M\'esz\'aros P., 1997, MNRAS, 288, L51
%
\bibitem{}
Yoshida A., \eal, 1997. In: Proc. of the 4-th Huntsville's GRB Symposium
(in press)
%

\end{thebibliography}
\end{document}